# Microstructural evolution during isothermal tempering of a highly supersaturated metastable austenitic state of a PM high-speed steel

S. Müller[a]*, S. Benito[a], H. Schaefer[a], Y. Li[b], L. Gertlowski[c,d], O. Schenk[d], U. Hagemann[e], C. Broeckmann[c,d], S. Weber[a]

[a]Ruhr University Bochum, Chair of Materials Technology, Bochum
[b]Ruhr University Bochum, Center for Interface-Dominated High Performance Materials, Bochum
[c]Institute of Applied Powder Metallurgy and Ceramics at RWTH Aachen e.V.
[d]RWTH Aachen University, Institute for Materials Applications in Mechanical Engineering, Aachen
[e]Interdisciplinary Center for Analytics on the Nanoscale, Duisburg

*Stefan Müller
Ruhr-University Bochum
Chair of Materials Technology
Building IC, Level 03, Room 313
Universitätsstraße 150
44801 Bochum
Germany
Stefan.mueller-f99@rub.de

## Abstract

This study investigates the microstructural evolution of the powder metallurgical high-speed steel PM HS 3-3-4 during isothermal holding within the transformation gap between pearlite and bainite formation. The focus is on the mechanism of carbide spheroidization and the proposed formation of a chemically graded carbide rim. The heat treatments were carried out in a quenching dilatometer, enabling precise control of the thermal cycles and isothermal holding conditions.

Spatially resolved analytic microscopy techniques, including scanning transmission electron microscopy (STEM), Auger electron spectroscopy (AES) and atom probe tomography (APT), image analysis and complementary thermodynamic simulations (MatCalc) were used to characterize the carbide evolution. Isothermal holding increased carbide circularity, carbide volume fraction, and hardness, while MatCalc calculations predict a strong thermodynamic driving force for additional carbide precipitation. The analytic techniques revealed a compositional gradient of approximately 15 nm at the carbide–matrix interface in the isothermally treated condition, compared with a sharp transition of only 2–3 nm in the directly quenched state. This nanometer-scale gradient provides strong evidence for epitaxial, interface-controlled carbide growth and the formation of a chemically graded carbide rim. The associated solute depletion of the surrounding austenite provides a plausible explanation for its reduced stability, refleclted by an increase in the martensite start temperature ($M_s$).

The combined results demonstrate that isothermal holding promotes interfacial chemical reorganization, offering new insights into carbide spheroidization and providing a basis for tailoring carbide morphology and the resulting mechanical properties through heat treatment.

# 1. Introduction

### 1.1 Motivation and Scientific Basis

High-quality tool steels are essential materials in modern manufacturing, as they enable the shaping of a wide range of materials and the production of components with diverse dimensions and geometries. Their application requires a combination of high strength, hardness and wear resistance while maintaining sufficient toughness [1]. This property profile is governed by a multiphase microstructure that can be tailored to specific application requirements through the manufacturing route, chemical composition, and thermomechanical treatment [2]. Within this material class, high-speed steels (HSS) are particularly relevant for highly loaded tooling applications. Their macroscopic properties arise from the combined contributions of the matrix and the different carbide populations, making control of the entire microstructure essential for optimizing material performance [1–3].

Powder metallurgical (PM) production of tool steels offers significant advantages over conventional ingot metallurgy, which are particularly relevant for highly alloyed grades such as HSS. The process, involving atomization of the melt into fine powder particles followed by hot isostatic pressing, yields a homogeneous, fine-grained microstructure with uniformly distributed carbides [4]. The PM production eliminates macroscopic segregation and largely prevents the formation of coarse primary carbides, resulting in improved toughness, enhanced wear resistance, and uniform mechanical properties across the entire cross-section of the material [5]. Moreover, PM technology allows for an extended alloying range, enabling higher concentrations of strong carbide formers without substantially compromising workability [6].

HSSs are characterized by a multiphase microstructure consisting primarily of a tempered α´-martensitic matrix reinforced by a dispersion of various types of carbides, such as MC, $M_6C$, and $M_2C$ [7, 8]. The distribution, size, and morphology of these carbides influence the wear resistance, toughness, and hot hardness of the steel. Heat treatment of HSS typically involves austenitizing, quenching, and multiple tempering cycles, each of which influences the resulting microstructure. The austenitizing temperature controls the dissolution of primary and secondary carbides, thereby determining the carbon and alloy content in the martensitic matrix. Subsequent quenching produces a high-hardness martensitic structure, which is tempered to relieve internal stresses, eliminate retained austenite, and precipitate tempering carbides, enhancing hot hardness and wear resistance. Precise optimization of these parameters allows for tailoring the balance between hardness, dimensional stability and toughness, ensuring optimal performance under the demanding conditions typical for cutting and forming operations.

A particular variant of heat treatment is quenching with an isothermal tempering stage, in which the workpiece, after austenitizing, is not cooled directly to room temperature, but first brought to a defined intermediate temperature within the gap between the bainitic and pearlitic transformations and kept there for a specified period. The transformation gap arises because, in this temperature range, the necessary diffusion rate for pearlite formation is already too low, while the nucleation energy for bainite has not yet been reached [9]. In the TTT diagram, this appears as a region without overlap between the start curves of the two microstructural constituents, such that the austenitic structure is kept in a metastable state as illustrated in the TTT diagram shown in Figure 2.

In highly alloyed tool steels, this gap is further widened because alloying elements such as Cr, Mo, or W delay the pearlite formation and shift the bainite start temperature $B_s$ to lower values [10]. This intermediate stage allows temperature gradients and quenching stress to be reduced without the formation of pearlite or bainite [11]. Such processes are particularly applied to minimize distortion and cracking in components with complex geometries, while preserving the required wear resistance. The isothermal holding stage within the transformation gap can not only reduce the risk of distortion and cracking but also enable targeted control of microstructural evolution, thereby enhancing macroscopic mechanical properties such as toughness and wear resistance [12]. Since macroscopic property changes are rooted in the microstructure, this represents a topic of scientific interest, especially as the field remains insufficiently investigated. Disch et al. demonstrated that an isothermal holding stage within the transformation gap of the powder-metallurgically produced high-speed steel PM HS 6-5-3-8 leads to a rounding of the contained undissolved $M_6C$ carbides [13]. Even short holding times of 160 s within the transformation gap result in an increase in carbide roundness, whereas prolonged holding promotes their uniform coarsening. The associated bonding of alloying elements, particulary carbon, in carbides leads to an elevation of the martensite start temperature. Consequently, the amount of retained austenite after quenching is reduced, which in turn results in increased hardness. The observed increase in carbide circularity after short tempering times in the transformation gap is attributed to epitaxial growth from the metastable, supersaturated austenite. Newly formed carbides are hypothesized to nucleate and grow on pre-existing carbides rather than by an additional nucleation and growth process. This secondary precipitation is expected to smooth irregular particle contours and thus to lead to an apparent spheroidization of the carbides, as reflected in the increased circularity values [13]. However, the proposed interface-controlled growth mechanism and the associated formation of a chemically graded rim around pre-existing MC carbides have so far remained hypothetical and have not yet been verified by spatially resolved experimental evidence.

### 1.2 Hypothesized mechanism behind carbide morphology changes

In the present study, the mechanism responsible for the spheroidization of carbides during isothermal tempering in the transformation gap is investigated. From the work of Disch et al., it is hypothesized that the carbide growth leading to the observed increase in circularity is accompanied by a chemical composition different from that of the original carbide. This compositional difference may manifest as a distinct “carbide rim” forming around the pre-existing carbides. The following section outlines this hypothesis in detail.

A possible explanation for the observed rounding of carbides is the interplay between restricted diffusion and epitaxial growth, which becomes effective within the transformation gap at the selected temperature.

During austenitization, thermodynamically stable, larger carbides grow at the expense of smaller, less stable ones via Ostwald ripening [14–16]. This growth is favored at grain boundaries because they provide fast diffusion pathways and thereby facilitate solute transport relative to the grain interior [17]. Consequently, carbide-forming alloying elements diffuse from the grain boundaries and their vicinity toward existing carbides, resulting in preferential growth along these boundaries [16, 17].

When the material is subsequently quenched rapidly to the holding temperature within the transformation gap (approx. 400–600 °C) [18], partially depleted grain boundaries and a supersaturated, metastable austenite remain in the vicinity of the existing carbides. Since the carbide phase is already present, no separate nucleation event is required. Instead, continued carbide growth is assumed to occur through epitaxial attachment at the existing carbide–austenite interface [19–21]. This growth is expected to be more pronounced on carbide surfaces facing the grain interior because the grain-boundary regions were depleted of carbide-forming elements during the preceding Ostwald ripening process. Preferential growth into the supersaturated austenitic matrix would consequently reduce the angularity of the carbides and result in an increasingly rounded morphology.

The carbon required for this precipitation can continue to diffuse along the grain boundaries toward the carbides even at lower temperatures, thereby promoting further growth [22]. Since the carbide grows outward into the supersaturated, metastable austenite, a carbide rim forms with a chemical composition similar to that of the surrounding austenite. Epitaxial growth is proposed to begin after only short holding times, providing a plausible explanation for the early onset of carbide rounding.

The mechanism described is schematically illustrated in Figure 1. The left panel shows the chemical gradients immediately after quenching from the austenitization temperature and the diffusion processes occurring during austenitization. As described, diffusion of all alloying elements takes place preferentially along grain boundaries, promoting carbide growth during austenitization. Rapid quenching then freezes the chemical composition between the carbide and the matrix.

The right panel illustrates the processes proposed to occur during isothermal holding within the transformation gap and the expected composition of the freshly formed carbide rim. At the holding temperature, carbon is assumed to be the only element with appreciable long-range mobility, whereas the substitutional alloying elements remain effectively immobile. The grain boundaries therefore act as fast diffusion pathways that transport carbon toward the pre-existing carbides. After reaching a carbide at a carbide–grain-boundary junction, carbon is proposed to redistribute along the carbide–austenite interface. The resulting carbon supply enables the existing carbide to grow epitaxially into the adjacent supersaturated austenite. Thus, the grain boundaries provide the pathway for carbon transport but do not necessarily determine the direction of interface migration. Because long-range redistribution of substitutional elements is strongly restricted, the newly formed rim is expected to largely inherit their local proportions from the surrounding austenite, resulting in a compositional gradient between the original carbide core and the matrix. Continued interfacial growth progressively reduces carbide angularity and produces a more rounded morphology.

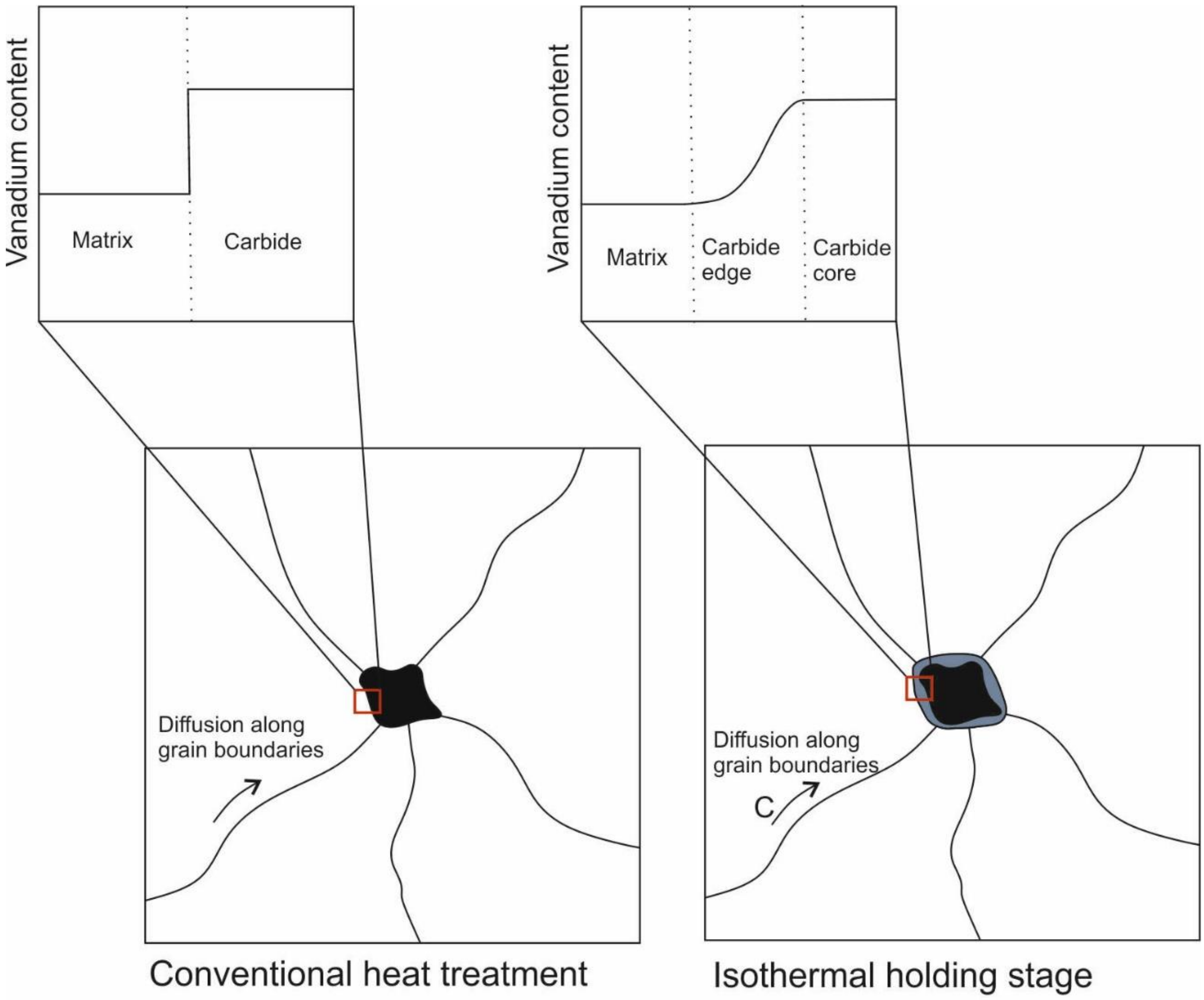

*Figure 1: Schematic illustration of the hypothesized mechanism behind carbide morphology changes.*

Several studies have demonstrated that such chemical inhomogeneities can emerge in carbides of tool steel [23, 24]. Furthermore, previous studies have shown that the distance between the nucleation sites, in this case the carbides, must be small enough to allow sufficient carbon to diffuse along the grain boundaries to the nucleation sites. [25, 26]. The chromium content plays a decisive role in determining diffusion rates and, consequently, the stability and morphology of carbides. The microstructural and chemical perspective presented here complements the previously proposed macroscopic hypothesis and provides a detailed explanation for the observed morphological evolution and early spheroidization of carbides [20, 25–27].

Assuming that carbide rounding during tempering within the transformation gap is accompanied by additional carbide growth, the width of the proposed carbide rim can be estimated. According to Disch et al. [13], the mean carbide size increases from approximately 0.6 µm² to 0.7–0.85 µm², depending on the holding time. In terms of radius, this corresponds to an increase of about 40–80 nm, suggesting the formation of a rim of comparable thickness.

### 1.3 Goal of this study

The objective of this study is to evaluate the hypothesis derived from the observations reported by Disch et al. regarding the mechanism responsible for carbide spheroidization during isothermal holding within the transformation gap. Based on this findings, the hypothesis investigated in the present work proposes that the observed increase in carbide circularity originates from epitaxial carbide growth, in which newly formed carbide regions develop on the surface of pre-existing carbides, resulting in the formation of a chemically graded carbide rim at the carbide–matrix interface.

To investigate this proposed mechanism, a sequential analytical approach was employed. In a first step, scanning transmission electron microscopy (S-TEM) was used to characterize the carbide morphology and the structural features of the carbide–matrix interfaces with nanoscale lateral resolution. The observations obtained from S-TEM provided the first key hints regarding the proposed interfacial growth mechanism. Subsequently, Auger electron spectroscopy (AES) was applied to obtain complementary nanoscale-depth information on the local chemical composition across the carbide interfaces and to identify indications of a potential compositional gradient associated with the proposed carbide rim. Finally, atom probe tomography (APT) was used to obtain three-dimensional compositional information at the near-atomic scale [28], providing further key evidence for the evaluation of the proposed chemical gradient at the carbide–matrix interface.

Each analytical technique therefore contributed essential key insights that, when considered together, enable a comprehensive assessment of the proposed carbide evolution mechanism. The experimental characterization was complemented by thermodynamic simulations and additional supporting analyses, which were used to interpret the experimental observations and evaluate the underlying driving forces for carbide evolution.

The manuscript follows this sequential analytical approach. First, the material, heat treatment conditions, resulting microstructure, and simulation results are presented to establish the basis for the subsequent investigations. The following section describes the STEM, AES, and APT analyses in that order, where the results from each technique progressively contribute to the evaluation of the proposed epitaxial carbide growth mechanism. Finally, the combined experimental and simulation results are discussed with respect to the validity of the hypothesis investigated in this work.

# 2. Material, heat treatment and microstructure

### 2.1 Material

In the present study, the powder metallurgically produced high-speed steel PM HS 3-3-4 (1.3377) was investigated. This material was selected because its hardened microstructure is comparatively simple, comprising a martensitic matrix and a single type of undissolved carbide, namely vanadium-rich MC carbides. The MC carbides provide sufficient contrast to the matrix, which is essential for the required image analysis [29].

Previous investigations identified these carbides as thermodynamically stable vanadium-rich MC carbides, originating from the transformation of metastable $M_2C$ carbides during powder metallurgical processing [29–31]. Owing to their high thermodynamic stability, only a limited fraction dissolves during austenitization, while the remaining particles act as stable constituents within the hardened microstructure.

The initial material was supplied by ZAPP AG in the form of a hot-isostatically pressed, hot-worked, and subsequently soft-annealed blank. The chemical composition was determined using optical emission spectrometry (OES). The obtained results, along with the nominal chemical composition, are presented in Table 1. The reported values represent the average of ten individual measurements. All values are given in weight percent (wt.-%).

**Table 1. Chemical composition of the investigated high-speed tool steel DIN EN 1.3377 in wt.-%.**

| | C | Si | Mn | P | S | Cr | Mo | V | W | N | Fe |
|---|---|---|---|---|---|---|---|---|---|---|---|
| Nominal composition | 1.5 | - | - | ≤0.03 | ≤0.03 | 4.0 | 2.5 | 4.0 | 2.5 | - | Bal. |
| Measured composition | 1.50 ±0.02 | 0.5 ±0.02 | 0.3 ±0.01 | 0.02 ±0.01 | 0.02 ±0.01 | 4.1 ±0.01 | 2.6 ±0.05 | 4.1 ±0.2 | 2.5 ±0.04 | 0.2 ±0.01 | Bal. |

### 2.2 Heat Treatment

All heat treatments were carried out using a quenching dilatometer of the type BÄHR DIL805 by TA Instruments. Experiments were conducted under vacuum atmosphere, with gaseous nitrogen used as the quenching medium. To perform the heat treatments in the quenching dilatometer, cylindrical samples with a diameter of 4 mm and a length of 10 mm were extracted from the initial material.

Since no time-temperature-transformation (TTT) diagrams are available for the investigated steel PM HS 3-3-4, initial heat treatments were performed to determine both the continuous cooling transformation (CCT) and the isothermal TTT diagrams. For the construction of the CCT diagram, all samples were austenitized at 1100 °C using a heating rate of 40 K/s. After a holding time of 30 minutes at 1100 °C, the samples were quenched to room temperature at various cooling rates. The applied cooling rates were represented by λ-values of 2, 8, 15, 25, 50, 120, and 180, where λ is defined as the cooling time from 800 to 500 °C in hectoseconds. By measuring the temperature-dependent change in sample length, the start and end temperatures of phase transformations could be determined.

Analogous to the CCT experiments, dilatometric tests were conducted to generate the isothermal TTT diagram. For this purpose, the samples were likewise austenitized at 1100 °C and held for 30 minutes. Instead of being quenched directly to room temperature, the cooling process was interrupted at various target temperatures, and the samples were held isothermally for 5 hours before being further cooled to room temperature. The selected holding temperatures ranged from 250 °C to 800 °C, varied in increments of 50 °C. As in the CCT tests, the phase transformation temperatures were determined using the tangent method based on the measured length changes. Based on the obtained data, both CCT and isothermal TTT diagrams were constructed using the tangent method.

Figure 2a presents the CCT diagram, while the corresponding isothermal TTT diagram is shown in Figure 2b. Furthermore, Figure 2a schematically illustrates the temperature profile of the reference samples, while Figure 2b depicts the temperature profile of the isothermally hold samples.

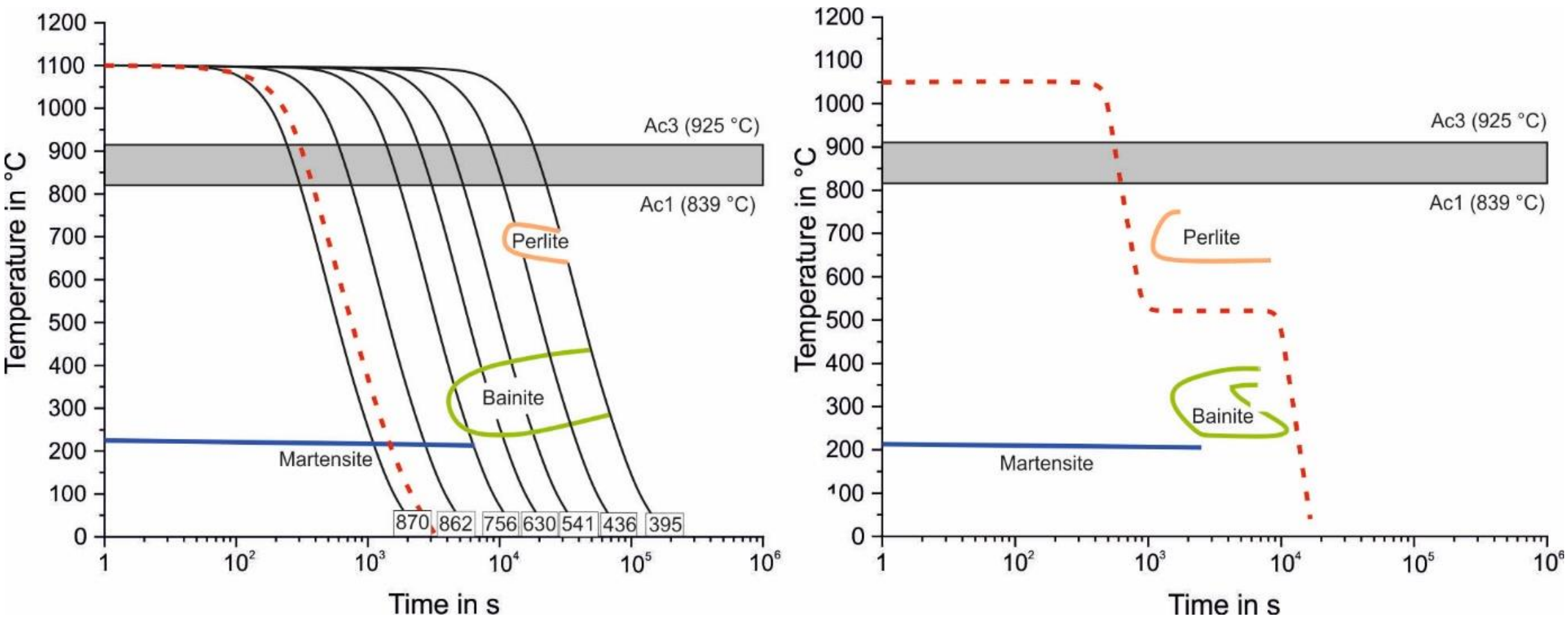

*Figure 2: Experimentally determined CCT and TTT diagrams for the investigated PM HS 3-3-4 steel.*

The resulting diagrams were used to define the heat treatment schedules for the investigated samples. A reference condition was established by austenitizing the material at 1100 °C followed by immediate quenching to room temperature without interruption, at a rate fast enough to avoid crossing any phase boundaries other than the α´-martensitic transformation. In addition, a modified heat treatment was performed, incorporating a holding stage during quenching. This step was applied at a temperature located in the center of the identified transformation gap between pearlite and bainite. The selected temperature was 520 °C, at which the samples were held for one hour before being quenched to room temperature.

After the heat treatments, the hardness of each sample was determined by performing five individual Vickers measurements (HV10), from which the mean value was obtained. The directly quenched reference samples exhibited a hardness of 877 ± 12 HV10, whereas the samples subjected to isothermal heat treatment at 520 °C showed a hardness of 899 ± 11 HV10.

## 2.3 Microstructure

### 2.3.1 Image analysis

Quantitative image analyses were carried out on both the directly quenched reference sample and the isothermally treated sample to obtain information on carbide morphology and volume fraction. For this purpose, the dilatometer specimens were sectioned at mid-length, and the cross-sections were metallographically prepared.

The specimens were mounted in electrically conductive epoxy resin and subsequently ground stepwise using SiC abrasive papers with mesh sizes ranging from 80 to 1000.

Polishing was performed in several stages using polycrystalline diamond suspensions, starting with 9 µm and finishing with 1 µm particle size.

The metallographically prepared samples were examined using scanning electron microscopy (SEM) in backscattered electron (BSE) and secondary electron (SE) mode for image analysis. A TESCAN Mira 3 microscope was employed, with a working distance of 10 mm and an accelerating voltage of 15 kV. For both the directly quenched reference condition and the isothermally treated condition, 15 images at 3000× and 15 images at 5000× magnification were acquired, corresponding to a total analyzed area of 68520 µm$^2$ per condition.

Image analysis was performed using the Fiji software package in combination with the Trainable Weka Segmentation plugin. This machine learning–based tool enables segmentation of the matrix and carbides by means of manually defined training regions [32]. To improve the comparability of the image analysis results, histogram matching was performed using the algorithm described by Burger in Digital Image Processing [33]. The corresponding Java™ implementation is available through the CorrectBleach plugin for ImageJ [34]. No additional image filtering or compression was applied during preprocessing to prevent any loss or distortion of the original image data. After segmentation, each image was binarized and converted into a two-phase representation, allowing each pixel to be classified as either matrix or carbide. Particles smaller than 120 nm were excluded from the analysis to avoid misinterpretation of isolated noise pixels as carbides. Based on this procedure, the carbide volume fraction was determined using polished samples analyzed under BSE contrast.

After the semantic segmentation, a distance-based watershed transform was used to perform the instance segmentation of the carbides. Finally, the circularity of each individual carbide as well as the mean circularity in each specimen were evaluated. For this purpose, images of etched samples acquired in SE contrast were used. Circularity serves as a measure of carbide roundness and was calculated according to the following equation:

$$Circularity = \frac{4\pi A}{p^2}$$

where *A* denotes the measured area and *p* the measured perimeter of the carbide. An ideal circle corresponds to a circularity value of 1, while decreasing values indicate increasingly elongated or irregular particle shapes [35].

Figure 3 presents representative micrographs obtained in BSE contrast for the directly quenched reference condition and the isothermally tempered comparison condition. In both images, a martensitic matrix containing finely dispersed MC-type carbides is visible. Figure 3 a) shows the directly quenched reference state, while Figure 3 b) displays the isothermally held condition. Below each micrograph, Figures 3 c) and d) show the corresponding binarized images obtained from image analysis, which were used for quantitative determination of the carbide volume fractions.

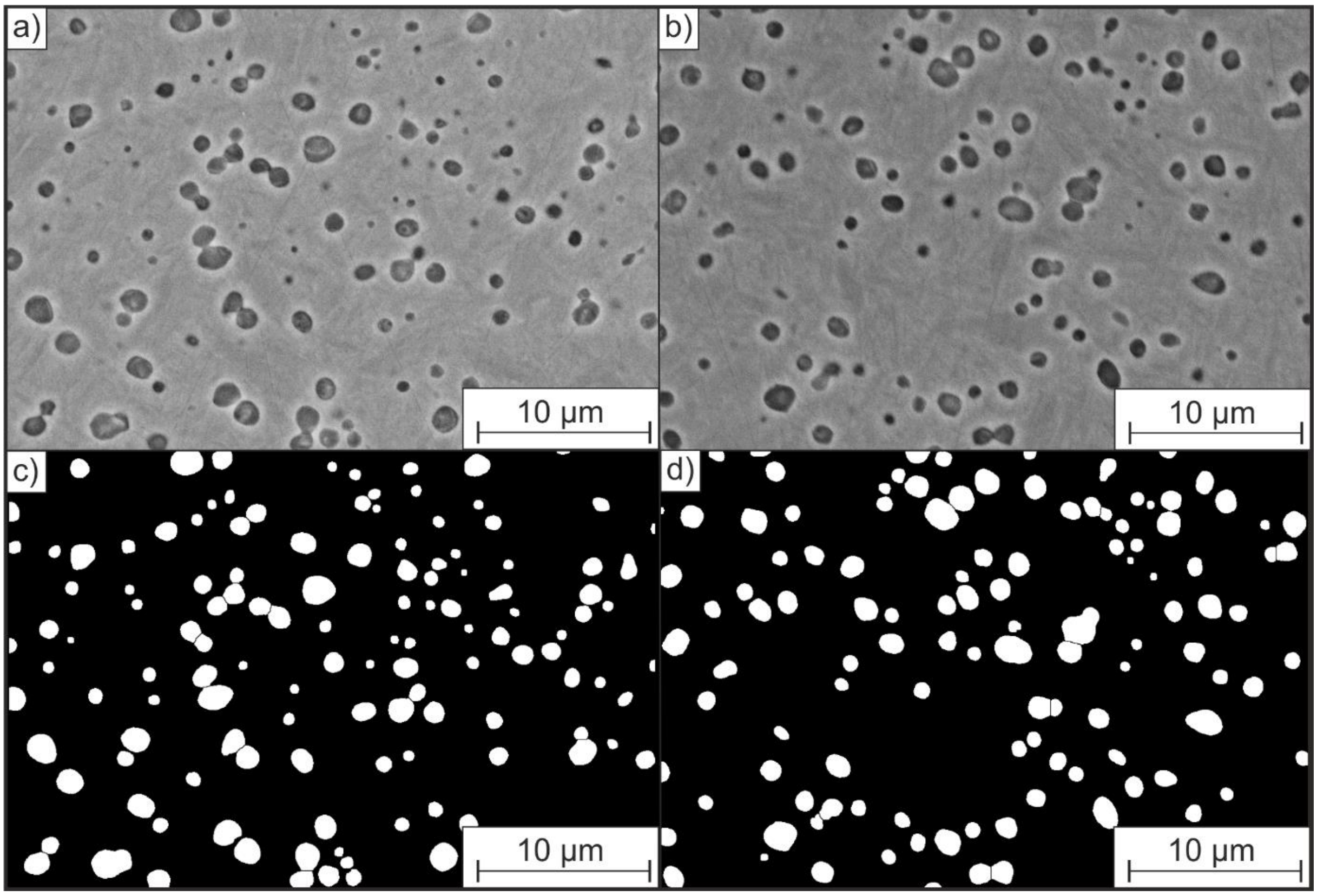

*Figure 3: SEM images of polished samples obtained in BSE contrast: (a) directly quenched reference condition and (b) isothermally held condition. (c) and (d) segmented microstructures of (a) and (b), respectively.*

The determined carbide volume fractions are summarized in Table 1. Approximately 1600 individual carbides were evaluated for each heat-treatment condition. The reported absolute measurement error was estimated stereologically using the standard discrete particle model to account for the total number of analyzed features and the geometric size variance of the individual particles across the analysed area. An increase in carbide volume fraction is observed after the isothermal holding step. Specifically, the carbide volume fraction increases from 8.3 ± 0.4 vol.% in the directly quenched reference samples to 9.9 ± 0.4 vol.% in the isothermally held samples.

*Tabelle 1: Carbide volume fraction in vol.%, determined by image analysis.*

| Carbide volume fraction without holding stage | Carbide volume fraction with holding stage |
|---|---|
| 8.3 ± 0.4 | 9.9 ± 0.4 |

In addition to the carbide volume fractions, the carbide morphology was characterized in both conditions using SE contrast acquired from Nital-etched specimens. The mean circularity of the MC carbides was calculated from binarized images obtained analogously to those shown in BSE contrast. Figure 4 a) again shows the directly quenched reference condition, and Figure 4 b) the isothermally held condition. The corresponding binarized images used for the determination of carbide morphology are displayed below in Figures 4 c) and d).

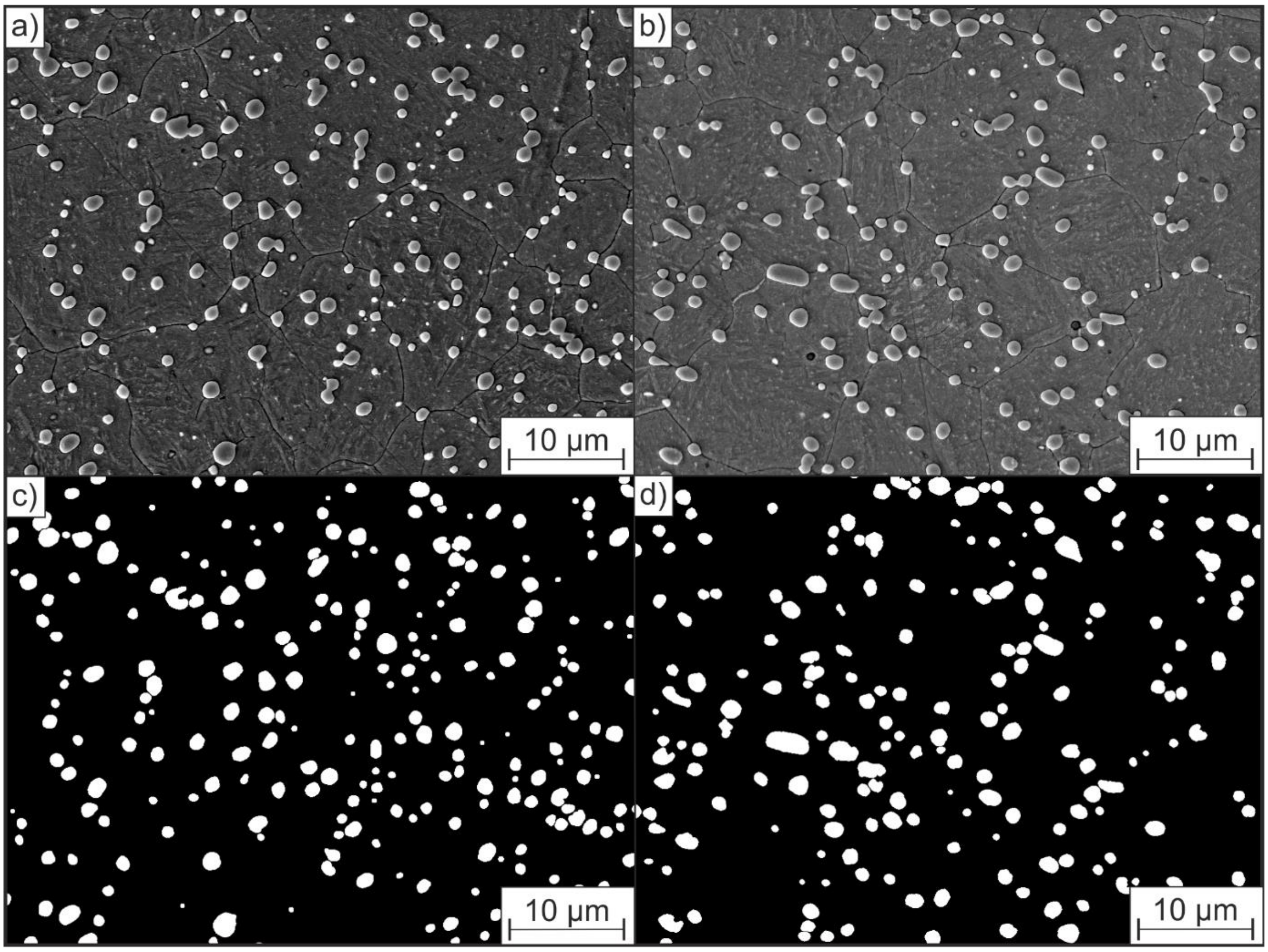

*Figure 4: SEM images of etched samples obtained in SE contrast: (a) directly quenched reference condition and (b) isothermally held condition. (c) and (d) segmented microstructures of (a) and (b), respectively.*

The measured mean circularity values of the MC carbides, including the associated standard deviations, are summarized in Table 2. An increase in carbide circularity was observed after the isothermal holding step. The mean circularity increased from 0.881

± 0.015 in the directly quenched reference condition to 0.914 ± 0.015 in the isothermally held condition. In addition, the mean area-equivalent carbide radius, determined on the unetched specimens, increased from 558 ± 180 nm in the directly quenched reference condition to 606 ± 192 nm after isothermal holding. The relatively large standard deviations reflect the broad carbide size distributions, indicating substantial size differences among individual carbides in both conditions. Although the comparison reveals an increase in the average size of the carbide population, it does not allow the extent of possible growth to be determined for individual carbides. Under the simplifying assumption that the observed increase results from uniform radial growth of the pre-existing carbides, the difference of 48 nm between the mean radii corresponds to an estimated average carbide-rim thickness of approximately 50 nm.

*Tabelle 2: Determined carbide circularity and mean radius.*

| Carbide circularity after direct quenching | Carbide circularity after isothermal holding | Mean carbide radius after direct quenching | Mean carbide radius after isothermal holding |
|---|---|---|---|
| 0.881 ± 0.015 | 0.914 ± 0.015 | 558 ± 180 nm | 606 ± 192 nm |

The increase in carbide circularity observed in the present study is less pronounced than that previously reported for isothermally tempered PM HS 6-5-3-8 [13]. This difference is consistent with the higher thermodynamic stability of MC carbides, which dissolve to a considerably smaller extent during austenitization compared to $M_6C$ carbides [29, 36, 37].

### 2.3.2 Simulation

To assess whether the hypothesized carbide growth during isothermal holding is thermodynamically and kinetically plausible, precipitation simulations were performed as a plausibility check of the proposed mechanism. In particular, the simulations were used to evaluate whether additional carbide precipitation and growth can occur under the selected holding conditions within the transformation gap. For this purpose, metastable microstructural states during the isothermal heat treatment were simulated using the software package MatCalc version 6.10, developed by MatCalc Engineering GmbH. MatCalc is a computational framework designed for the simulation of phase transformations, precipitation kinetics, and microstructural evolution in metallic materials [38]. The calculations are based on the thermodynamic extremal principle formulated by Lars Onsager, which describes the interactions of systems under non-equilibrium thermodynamic conditions [39, 40]. To account for concurrent material processes, MatCalc combines multiple fundamental models, enabling the consideration of coupled interactions occurring simultaneously within the material. The thermodynamic basis of the simulations is the CALPHAD approach in combination with associated databases. These experimentally validated databases provide thermodynamic, chemical, and physical quantities, including Gibbs energies, molar volumes, and diffusion coefficients.

To perform the simulations, the global chemical composition of the PM HS 3-3-4 steel determined by OES (Table 1) was first implemented into the model. For the simulation of the isothermal heat treatment, the actual quenching process in the dilatometer occurred sufficiently rapidly to preserve the carbide microstructure present at the end of the austenitization step [29]. Consequently, the experimentally measured carbide fractions of the directly quenched reference condition were used as input parameters for the subsequent simulation of the isothermal treatment. The corresponding carbide fractions determined by quantitative image analysis were incorporated into the model and served as the initial condition for the simulation of the isothermal holding stage. Thus, the starting point of the simulation was defined by the measured global chemical composition together with the carbide distribution obtained from image analysis. Based on these initial conditions, an isothermal holding treatment at 520 °C for 1 h was simulated. The resulting carbide fractions were evaluated and plotted as a function of the holding time at 520 °C. Figure 5 presents the temporal evolution of the carbide volume fraction.

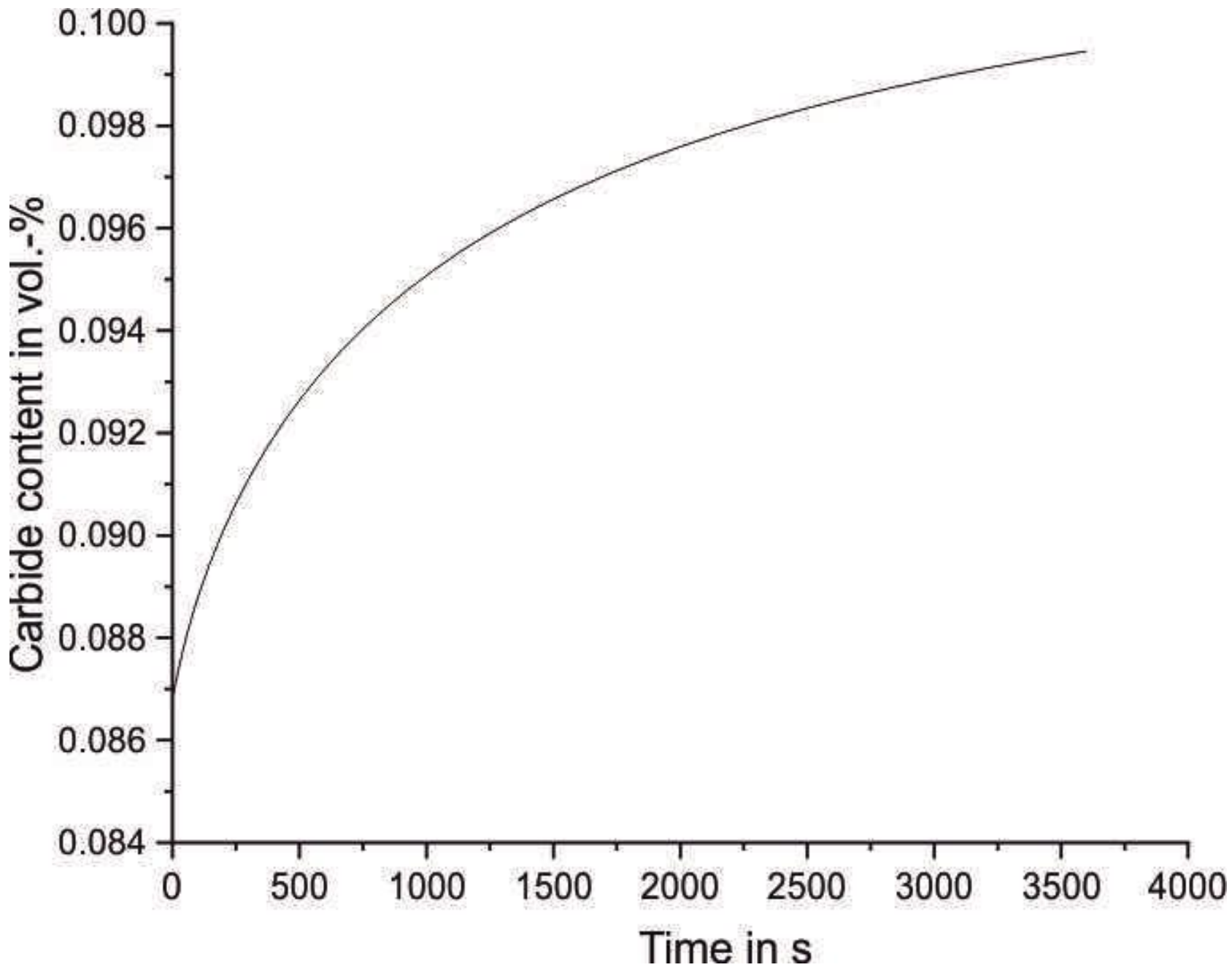

*Figure 5: Simulated carbide volume fraction evolution during tempering in the transformation gap.*

The MatCalc simulations indicate a pronounced thermodynamic driving force for carbide precipitation during isothermal holding within the transformation gap. Accordingly, the calculated carbide volume fraction increases from 8.5 vol.% to 9.9 vol.%. This result supports the experimentally observed increase in carbide content and demonstrates that carbide formation is thermodynamically favored under the applied holding conditions. However, the simulation does not reveal whether the additional carbide fraction results from the nucleation of new carbides, the growth of pre-existing carbides, or a combination of both mechanisms.

The interpretation of the results is limited by the underlying mean-field approach, in which spatial information is not explicitly considered. Consequently, local concentration gradients, solute redistribution in the vicinity of existing carbides, particle interaction zones, and the influence of grain boundaries cannot be resolved. Possible local variations in the thermodynamic driving force that may promote preferential growth or nucleation at existing carbide–matrix interfaces therefore remain inaccessible.

From a microstructural perspective, epitaxial growth at existing carbide interfaces appears plausible. Following quenching from the austenitization temperature to the isothermal holding temperature, the microstructure consists predominantly of metastable austenite with a relatively low dislocation density. As a result, only a limited number of energetically favorable intragranular nucleation sites are expected [41, 42]. Existing carbide-matrix interfaces, on the other hand, provide energetically favorable sites for continued carbide growth because the carbide phase is already present and no separate nucleation barrier must be overcome [43]. Minimizing the interfacial-energy contribution to the total free energy provides an additional energetic argument for the growth of pre-existing carbides. Carbide-forming elements can therefore attach directly to the existing carbide lattice, enabling interface-controlled growth and, where sufficient crystallographic compatibility exists, potentially epitaxial growth [44].

Nevertheless, these mechanisms are not explicitly represented in the current simulation framework. Newly formed carbides and growth of existing carbides are treated statistically and cannot be distinguished. Consequently, the simulations neither confirm nor refute the formation of a carbide rim resulting from preferential heterogeneous nucleation at existing carbides.

Overall, the MatCalc simulations demonstrate the existence of a significant thermodynamic driving force for carbide precipitation during isothermal holding within the transformation gap and confirm the increase of carbide volume fraction. However, they do not provide sufficient spatial or mechanistic information to identify whether this increase originates from carbide nucleation, carbide growth, or both. Therefore, nanoscale investigations of the carbide-matrix interfaces were performed to identify potential compositional gradients associated with rim formation during isothermal tempering.

# 3. Analytic investigation of the proposed carbide epitaxial growth

### 3.1 Observation of the carbide interfaces with nanoscale lateral resolution

Scanning transmission electron microscopy (STEM) investigations were carried out to analyze the chemical distribution across individual carbide-matrix interfaces. For this purpose, a thin platinum layer was first deposited onto the sample surface above the region of interest using the gas-injection-system of an FEI Helios NanoLab G3 CX Dual-Beam FIB-SEM. Subsequently, lamellae were extracted by focused ion beam (FIB) milling and progressively thinned to a final thickness of approximately 50 nm using stepwise reduced ion beam currents. STEM measurements were performed at an acceleration voltage of 200 keV. Elemental analysis was conducted using energy-dispersive X-ray spectroscopy (EDS) in line-scan mode across selected MC precipitates and the surrounding martensitic matrix. The effective electron probe diameter during EDS acquisition was approximately 1–2 nm, enabling high lateral resolution across the carbide-matrix interface. Line scans were performed perpendicular to the carbide boundary to assess the spatial variation of vanadium and iron signals and to evaluate potential signal overlap between the MC precipitates and the martensitic matrix. The measurements were performed on three carbides from the directly quenched reference state and on three carbides from the isothermally tempered samples. In both cases, the carbides were randomly selected.

Figure 6 shows the three results of the STEM investigations performed on a carbide in the directly quenched reference state, while Figure 7 presents the three STEM results obtained from a carbide in the isothermally tempered state. In both cases, the ratio of the measured intensities of vanadium and iron is shown. Vanadium and iron were selected because they represent the dominant elemental contrast between the V-rich MC carbide and the Fe-rich matrix, respectively, thereby providing a sensitive measure of the chemical transition across the carbide–matrix interface. In the reference state, a compositional gradient with a spatial extent on the order of several tens of nanometers can be observed. Similar gradients are also detected in the isothermally tempered state, with measured spatial extents generally ranging from a few tens of nanometers up to approximately 100 nm. Overall, the observed gradients in both conditions are of the same order of magnitude, and no clear distinction in gradient width can be established based on the present measurements.

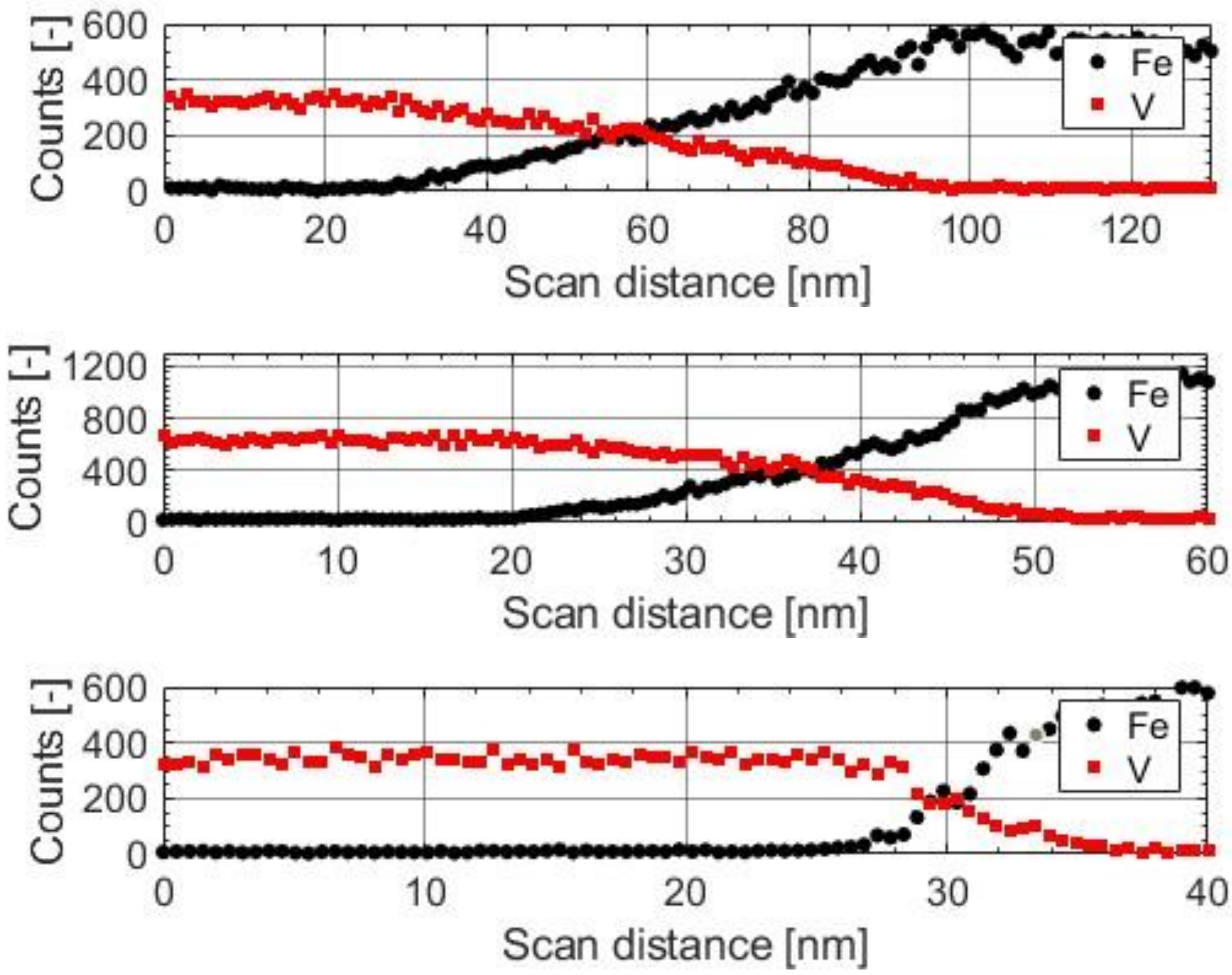

*Figure 6: Fe and V profiles at the interface between the carbide (left) and the matrix of the reference state (right), as determined by STEM.*

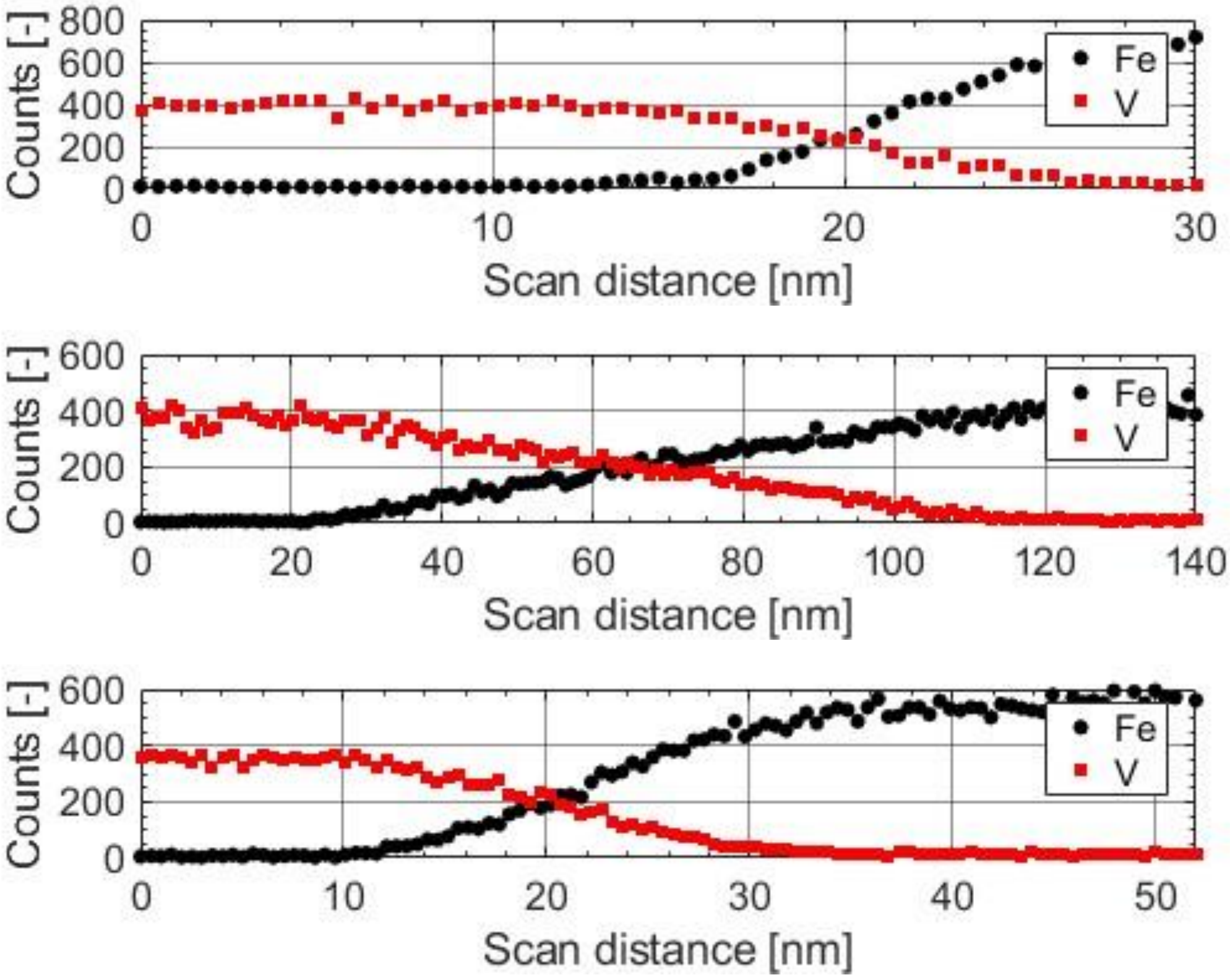

*Figure 7: Fe and V profiles at the interface between the carbide (left) and the matrix (right) of the isothermal tempered state, as determined by STEM.*

To validate the STEM measurements, both Monte Carlo simulations and geometrical calculations were performed. The Monte Carlo simulations were used to investigate the interaction volume of the electron beam with the microstructure, whereas MATLAB calculations were conducted to evaluate the influence of carbide geometry on the measured signal. Monte Carlo simulations of electron-matter interactions were performed using the CASINO software package (Monte CArlo SImulation of electroN trajectory in sOlids) to support the interpretation of the experimental STEM-EDS results [45]. The simulations were used to estimate the electron scattering behavior, the effective interaction volume, and the spatial distribution of X-ray generation within the carbide and the surrounding matrix. The geometrical model consisted of a specimen of thickness 50 nm, corresponding to those used in the experiments. A perfectly sharp transition parallel to the electron beam was used as the carbide-matrix interface. The incident electron beam energy was set to 200 keV, corresponding to the experimental STEM conditions, and the beam was assumed to be normally incident on the specimen surface. The simulations were performed with a linearly moving electron source across the phase boundary to evaluate the extent to which X-rays generated in the surrounding Fe matrix contribute to the measured EDS signal when the electron probe is positioned within the MC precipitate.

Figure 8a shows the results of the CASINO simulation X-ray intensity for vanadium as a function of position. Here, the zero position indicates the interface. The simulation demonstrates that the spatial resolution of the STEM-EDS measurements is sufficient to reliably detect the assumed carbide rim with a width of 50–80 nm. Based on the simulated electron–matter interactions, the expected measurement uncertainty is approximately 4 nm, which is significantly smaller than the characteristic width of the investigated compositional gradients.

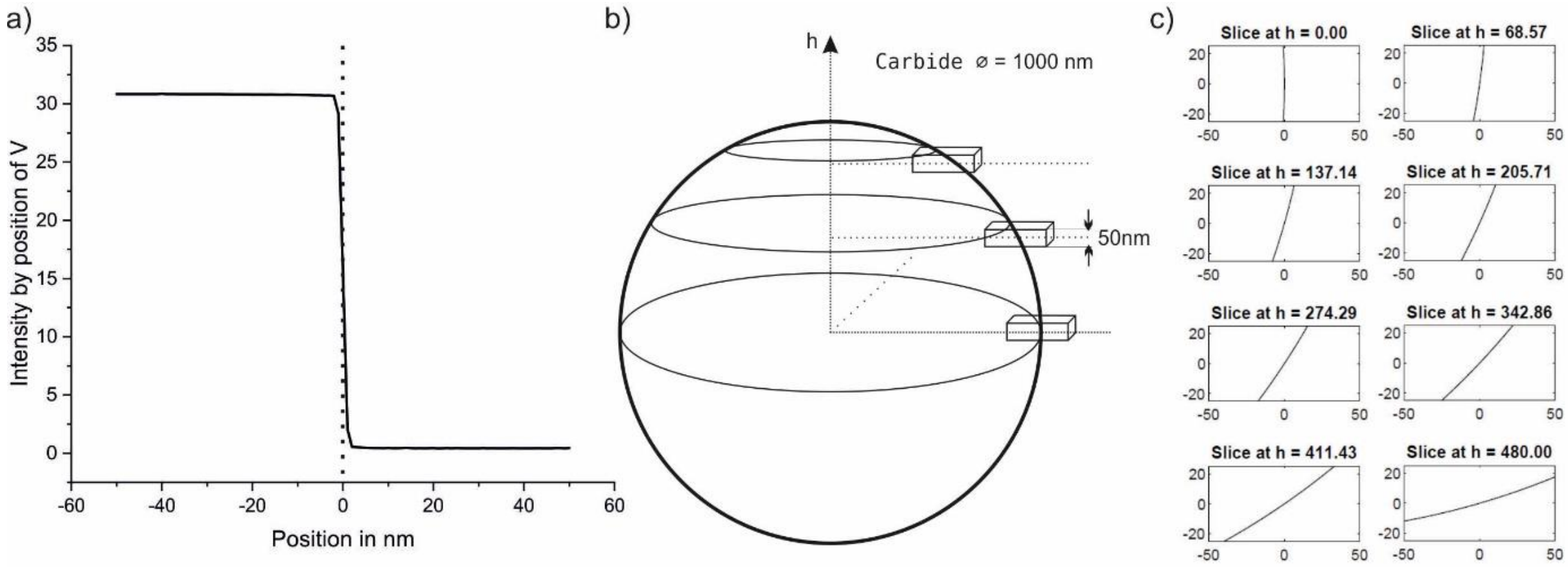

*Figure 8: Assessment of spatial broadening in STEM–EDS measurements using CASINO simulations and MATLAB-based geometric calculations: (a) simulated vanadium X-ray intensity profile across an ideally sharp carbide–matrix interface located at position zero; (b) positions of the investigated cross sections through a spherical MC carbide with a diameter of 1000 nm, where h = 0 corresponds to the particle center; and (c) influence of section height h on the calculated contributions of carbide and Fe matrix to the sampled volume. A specimen thickness of 50 nm was used in both models.*

Since the Monte Carlo simulations assume a perfectly sharp and straight interface, additional considerations should account for the carbide size and shape. To that end, MC precipitates were modeled as ideal spheres with a diameter of 1000 nm and cross-sectional planes at different heights, h, within the carbide were considered. These calculations, performed in MATLAB, provide insights into the effect of the position along the vertical axis of the particle on the measured signal (Figure 8). The specimen thickness was set to 50 nm, consistent with the value used in the CASINO simulations. Figure 8 b illustrates the location of the respective cross-sectional planes within the carbide, where h = 0 corresponds to a section passing through the center of the spherical precipitate. The results of the geometric considerations are presented in Figure 8 c. In the central region of the carbide (h = 0–68 nm), the measurement volume is located almost entirely within the precipitate, resulting in a negligible contribution from the surrounding Fe matrix. With increasing section height (h = 137–342 nm), the fraction of simultaneously sampled matrix material increases due to the curvature of the spherical geometry. In regions close to the carbide boundary (h = 411–480 nm), the contribution of the Fe matrix becomes significant, as only a small fraction of the measurement volume remains within the carbide.

The STEM-EDX measurements provide the first direct nanoscale indication of a gradual compositional transition across the carbide–matrix interface. The complementary CASINO simulations and geometrical calculations further show that part of the measured broadening may arise from electron scattering and the three-dimensional carbide geometry. Nevertheless, the experimentally observed concentration profiles cannot be attributed unambiguously to these effects alone. To further evaluate whether a chemically modified interfacial region is present, an independent technique with substantially higher surface sensitivity was therefore employed. Auger electron spectroscopy (AES) was selected to investigate the chemical composition within the near-surface region of the carbide–matrix interface, probing a depth of approximately 3–5 nm [46].

### 3.2 Nanometer-depth chemical analysis of the carbide interface

Auger electron spectroscopy (AES) measurements were performed using a PHI 710 Scanning Auger Nanoprobe (ULVAC-PHI, Japan) to further investigate the chemical distribution across individual carbide–matrix interfaces. AES is a highly surface-sensitive technique in which a focused primary electron beam excites element-specific Auger electrons. Owing to the short escape depth of these electrons, the detected signal originates predominantly from the near-surface region, thereby substantially reducing signal averaging through the specimen thickness compared to STEM-EDS [46, 47]. The AES measurements were deliberately focused on the isothermally tempered condition because this state is expected, according to the proposed hypothesis, to exhibit the chemically modified carbide-rim region. The aim of this intermediate analytical step was therefore not to reproduce the complete comparison between both heat-treatment conditions, but to determine whether compositional gradients could be detected at carbide–matrix interfaces in the condition in which additional carbide growth is expected.

Line scans were acquired across selected carbides with a step size of 19 nm, providing high spatial resolution of vanadium and iron signals at the nanoscale. The primary electron beam energy was 3kV and the beam current 1nA. Measurements focused on regions encompassing entire MC precipitates and adjacent matrix areas to capture compositional gradients and interface characteristics.

A single scan was performed along a line that was deliberately positioned in such way that two carbides - and thus four interfaces - were examined for the isothermally tempered sample. As depicted in Figure 9 the first transition from the matrix into the carbide, two measurement steps, corresponding to approximately 40 nm, are required to reach a constant iron-vanadium ratio in the carbide (left panel in Figure 9). In contrast, for transitions two and three (exit from the first carbide and entry into the second carbide, central panel in Figure 9), a constant ratio is already found after only a single measurement step, indicating that no compositional gradient exists within the resolution of the measurement setup. For the final transition (right panel in Figure 9) from the carbide back into the matrix, roughly three measurement steps, corresponding to approximately 60 nm, are required to reach a constant metal ratio, again indicating the presence of the hypothesized compositional gradient.

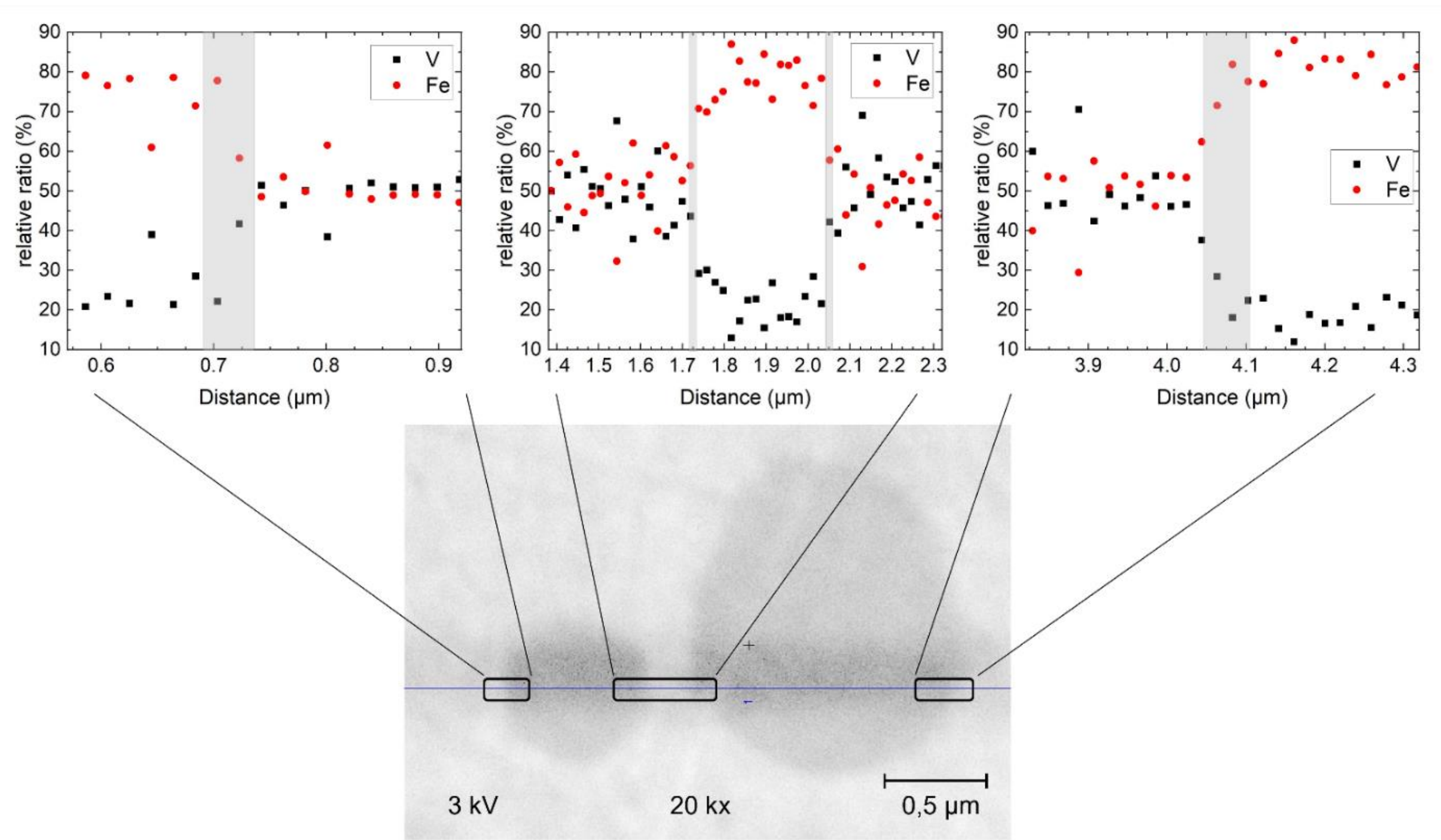

*Figure 9: Results of AES-measurements at the isothermal tempered state.*

Consequently, compositional gradients were detected at two of the four investigated interfaces, whereas the remaining two interfaces exhibited transitions that were sharp within the spatial resolution of the measurement. Notably, the two sharp transitions correspond to the closely spaced, opposing interfaces between two neighboring carbides. One possible explanation is that the small matrix region separating these carbides represents a limited local solute reservoir. Overlapping diffusion fields or preceding solute redistribution toward both adjacent carbides could locally reduce the supersaturation available for further carbide growth. Consequently, the driving force or available solute supply for additional precipitation at these opposing interfaces may be smaller than at interfaces facing a larger volume of surrounding matrix.

Compared with STEM-EDS, AES offers substantially higher surface sensitivity and is therefore less affected by signal averaging originating from the interaction volume of the incident electron beam [47]. The observed gradients of approximately 40–60 nm are consistent with the dimensions expected for the proposed carbide rim and therefore provide additional indications that chemical modifications may occur at selected carbide–matrix interfaces during isothermal holding. However, the observed behaviour is not consistent across all investigated interfaces. Furthermore, the measurements were constrained by a trade-off between spatial resolution and signal intensity. Consequently, only two to three measurement points are located within the expected transition zone, preventing a reliable characterization of the concentration profile. The AES measurements therefore provide further indications of a possible compositional gradient but do not allow an unambiguous assessment of its shape or width.

To resolve the chemical distribution across the carbide–matrix interface without the limitations imposed by the AES step size and to quantify the gradient with near-atomic spatial resolution, atom probe tomography (APT) was employed as the final analytical technique. In contrast to STEM-EDS and AES, APT enables three-dimensional compositional analysis on the atomic scale and therefore provides the spatial resolution required to directly evaluate the proposed carbide-rim hypothesis [48].

### 3.3 Atomic-scale 3D quantification of the interfaces

APT specimens were prepared using a dual-beam focused ion beam/scanning electron microscope (FIB/SEM; FEI Helios G4 CX) following the general procedure described in [49]. Regions of interest were first protected by depositing a Pt layer. Site-specific sections containing a carbide–matrix interface were then extracted using an in situ lift-out procedure and mounted on suitable support posts. Carbides that were visible in the two-dimensional section of the metallographic sample were targeted at their edge regions to position the carbide–matrix interface within the final analysis volume. The mounted sections were subsequently shaped by annular FIB milling until a tip diameter of less than 50 nm was obtained. Additional specimens containing only matrix material were prepared using the same procedure to investigate localized chemical variations outside the carbide–matrix interfaces.

The APT measurements were performed in voltage mode at 70 K, a pulse fraction of 15%, a pulse repetition rate of 200 kHz, and a detection rate of 0.005 atoms per pulse on a local electrode atom probe (LEAP 5000 XR™, Cameca Instruments) for the reference samples. For the isothermally tempered samples, laser pulsing mode was used at a base temperature of 65 K and a laser pulse energy of 50 pJ, while the pulse repetition rate and target detection rate were kept unchanged. Laser pulsing was employed to improve the APT analysis success rate and data yield because the samples were more susceptible to premature fracture under voltage pulsing conditions. The APT datasets were reconstructed and analyzed using AP Suite software.

Figure 10 presents the APT results for the reference and isothermally tempered states. The proximity histograms in Figure 10a and b show the concentration profiles normal to the carbide-matrix interfaces, calculated relative to an iso-concentration surface of 20 at.% V. Figure 10c shows a representative reconstruction of the reference sample, in which part of an MC carbide is captured at the tip apex and identified by the enrichment of V atoms shown in green. Figure 10d shows the corresponding reconstruction of the isothermally tempered state, with the carbide-matrix interface located in the lower part of the reconstructed volume.

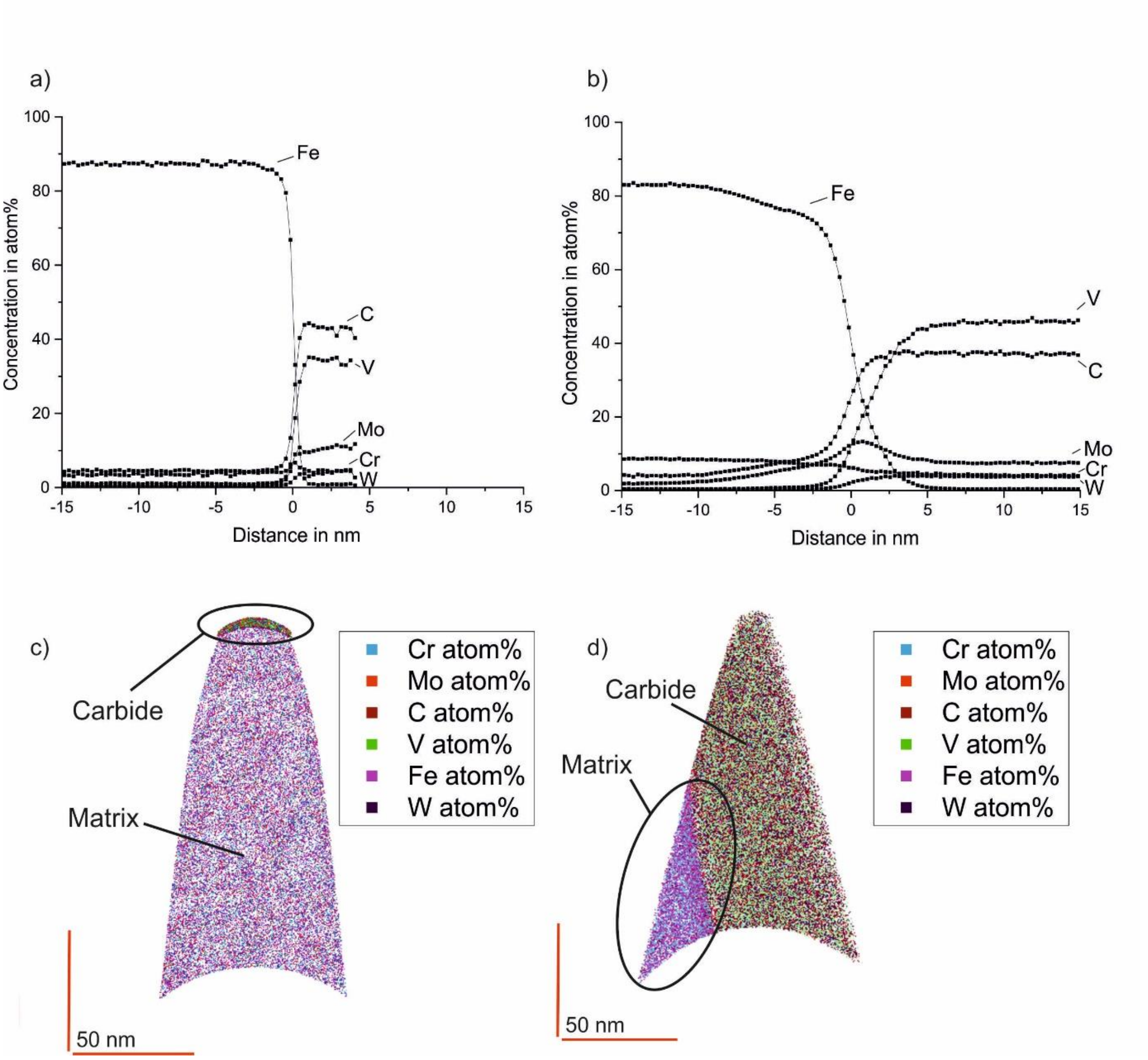

*Figure 10: APT analysis of the carbide-matrix interface in the reference and isothermally tempered conditions. (a,b) Proximity histograms across the carbide-matrix interface for (a) the reference condition and (b) the isothermally tempered condition. (c,d) Representative three-dimensional APT reconstructions showing the carbide and surrounding matrix in (c) the reference condition and (d) the isothermally tempered condition. In the reference specimen, the carbide is located near the apex of the reconstructed volume, whereas in the isothermally tempered specimen the carbide-matrix interface extends approximately parallel to the analysis direction.*

The APT measurements reveal a substantially broader carbide–matrix interface after isothermal holding than in the directly quenched reference condition. In contrast to the chemically sharp interface observed in the reference state, the isothermal held condition exhibits a systematic and element-dependent compositional transition extending over approximately 10–15 nm. The consistent behaviour of the Fe, C, V and Mo concentration profiles strongly indicates that the interface undergoes chemical reorganization during isothermal holding.

The quantitative width of the measured gradient should nevertheless be interpreted with appropriate caution because the two datasets were acquired under different experimental and geometrical conditions. The dissimilar evaporation behaviour of the carbide and martensitic matrix can give rise to local magnification effects and ion-trajectory aberrations, which may artificially broaden reconstructed interfaces [50]. Furthermore, the interface in the reference specimen was oriented approximately perpendicular to the analysis direction, whereas the interface in the isothermally held specimen was nearly parallel to the tip axis. The latter geometry is probed predominantly through the lower lateral spatial resolution of APT and is therefore more susceptible to trajectory-induced broadening. Additional differences may arise from the use of voltage pulsing for the reference condition and laser pulsing for the isothermally held condition.

Nevertheless, the systematic and element-dependent concentration gradients observed after tempering suggest that the broadening is unlikely to be solely an APT artefact. During carbide growth, substitutional elements must partition across the advancing carbide-matrix interface, and their comparatively slow and element-dependent redistribution may lead to local enrichment or depletion and incomplete compositional equilibration near the growth front. The measured interface width is therefore considered to contain both an APT-related contribution and a genuine chemically graded region associated with isothermal holding-induced carbide growth [50].

Besides the carbide-matrix interfaces, localized carbon-enriched regions were observed within the martensitic matrix of both heat-treatment conditions. These are shown in Figure 11 a) for a tip containing only matrix from the reference state and in Figure 11 b) for a corresponding matrix-only tip from the isothermally aged state. These regions exhibit a characteristic size on the order of a few nanometers.

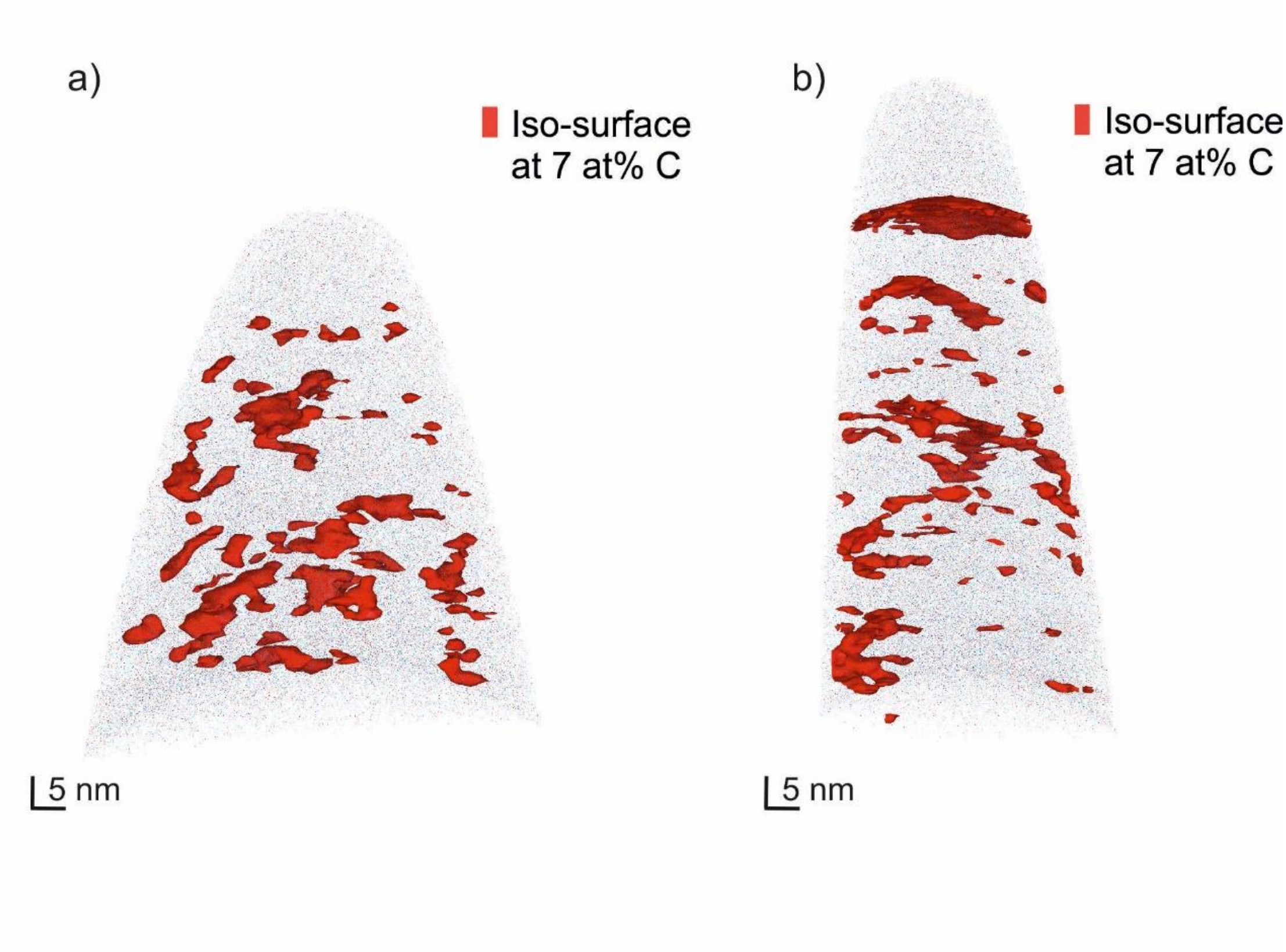

*Figure 11: Carbon distribution in the matrix measured by APT. 3D APT reconstructions of the matrix in (a) the reference condition and (b) the isothermally held condition. Carbon-rich regions are visualized using iso-concentration surfaces plotted at 7 at.% C.*

The localized carbon enrichments observed within the matrix are present in both heat-treatment conditions and therefore cannot be directly related to the interfacial compositional gradient. Their origin and possible significance for carbide evolution are discussed together with the remaining experimental observations in the following chapter.

# 4. Evaluation of the hypothesized carbide evolution

The experimental observations obtained in this study provide a consistent sequence of evidence for evaluating the proposed mechanism of carbide evolution during isothermal holding within the transformation gap. The transformation gap in the TTT and CCT diagram of high-speed steels is conventionally interpreted as a temperature interval in which neither pearlitic nor bainitic transformation occurs on experimentally relevant timescales. However, the absence of a detectable bulk phase transformation does not necessarily imply that the supersaturated austenite and the carbide population remain chemically or morphologically unchanged. Isothermal salt-bath treatments of several highly alloyed tool steels, including PM HS 6-5-2C, have previously shown that carbide size, microstructure, and hardness can change during intermediate-temperature holding [51]. More specifically, Disch et al. [13] demonstrated for PM HS 6-5-3-8 that isothermal tempering between 450 and 600 °C increases $M_6C$ carbide circularity and promotes further carbide growth. These results challenge the interpretation of the transformation gap as a microstructurally inactive temperature interval.

The importance of carbide reactions for the metastable state of PM HS 3-3-4 is also supported in a previous work by some of the authors of this article [29]. For the same steel grade as that investigated here, their combined dilatometric and MatCalc study showed that carbide dissolution and the associated redistribution of alloying elements during short-term heat treatment determine the chemical composition of austenite and influence the subsequent martensite-start temperature. Although that study focused primarily on austenitization rather than isothermal holding within the transformation gap, it demonstrates that even comparatively short heat-treatment stages can measurably alter the carbide–matrix equilibrium of PM HS 3-3-4.

Carbide precipitation at temperatures around 500–600 °C is well established in the broader literature on high-speed steels. Stiller et al. [52] used analytical electron microscopy and atom-probe field-ion microscopy to show that extremely fine $M_2C$ precipitates form during tempering of powder-metallurgical ASP 60 at 560 °C. Fischmeister et al. [53] subsequently demonstrated by atom-probe analysis that secondary-hardening precipitates in high-speed steels are compositionally complex and may contain Mo, Cr, V, W, and Fe. Rong et al. [54] further showed that the precipitation sequence, phase selection, and chemical composition of MC and $M_2C$ carbides depend strongly on the overall alloy composition. Pippel et al. [55] identified primary MC and $M_6C$ carbides as well as nanometer-sized V- and C-rich secondary carbides in HS 6-5-2-5 and observed coherent lattice-plane transitions between some of the nanoscale precipitates and the matrix. These studies primarily concern precipitation from a defect-rich α-martensite during tempering, whereas the present isothermal holding step is applied before final quenching, when the matrix remains predominantly metastable austenitic. The available nucleation sites and diffusion conditions are therefore different. Nevertheless, the literature establishes that appreciable redistribution of carbon and substitutional carbide-forming elements is possible in high-speed steels within the investigated temperature range.

A relevant conceptual framework for discussing interface-assisted carbide growth is the divorced eutectoid transformation (DET). In the classical DET described by Verhoeven and Gibson [26, 56], pre-existing cementite particles act as sinks for carbon while an austenite–ferrite transformation front advances. Carbon is incorporated into existing particles rather than forming cooperatively growing cementite lamellae, resulting in a dispersion of spheroidal carbides instead of pearlite. Phase-field simulations by Ankit et al. [57] further showed that the transition between cooperative pearlitic growth and non-cooperative divorced growth depends strongly on the undercooling and spacing of the pre-existing carbide particles. The present transformation cannot be classified as a classical DET because no migrating austenite–ferrite transformation front was detected and the investigated particles are MC carbides rather than cementite. Nevertheless, the DET concept provides an important mechanistic analogy. In both cases, pre-existing carbides may act as energetically favorable sinks at which the supersaturation of the surrounding matrix is relieved. This enables carbide growth without the cooperative formation of a lamellar transformation product. The present mechanism may therefore be described as DET-like or as non-cooperative, interface-assisted carbide growth.

The quantitative image analysis provides the first direct indication of carbide evolution in the present material. Both the carbide volume fraction and the mean carbide circularity increase after isothermal holding. The increase in circularity indicates a reduction in interfacial irregularity and is consistent with a tendency to reduce the carbide–matrix interfacial energy. Disch et al. [13] observed a comparable, although more pronounced, rounding of $M_6C$ carbides in PM HS 6-5-3-8. The present results suggest that an analogous process may occur at the MC carbides in PM HS 3-3-4. However, the smaller increase in circularity compared with that previously reported for $M_6C$ carbides [13] is consistent with the different thermodynamic stability and chemical constitution of the carbide phases. MC carbides in high-speed steels are mostly V-rich and comparatively stable, whereas $M_6C$ carbides contain larger fractions of Mo and W and may respond differently to changes in the chemical potential of the surrounding matrix [54, 55]. Furthermore, the lower austenitization temperature applied in the present study (1100 °C instead of 1150 °C) reduces the amount of alloying elements dissolved into the matrix and consequently the driving force for the proposed additional carbide precipitation during isothermal holding. The morphological evolution observed in different high-speed steel grades can therefore not be expected to be quantitatively identical. In addition to the changes in volume fraction and circularity, the mean equivalent carbide radius increases from 558 ± 180 to 606 ± 192 nm during isothermal holding. This value is of the same order as the carbide-rim thickness of approximately 40–80 nm estimated from the morphological observations of Disch et al. [13]. However, this estimate must be interpreted cautiously. The carbide population exhibits a broad size distribution. The calculation further assumes that comparable particle populations and sectioning planes are sampled in both conditions and that the mean-radius difference results exclusively from uniform interfacial growth. Precipitation of additional particles, selective dissolution of small carbides, coarsening, and stereological effects may also shift the measured mean radius. The value of 48 nm should therefore be regarded as an order-of-magnitude estimate of the apparent radial growth rather than as a direct measurement of the carbide-rim thickness.

The MatCalc simulations demonstrate a thermodynamic driving force for continued carbide formation during isothermal holding and predict an increase in carbide volume fraction from 8.5 vol.% to 9.9 vol.%. Classical late-stage Ostwald ripening primarily involves the redistribution of material between particles at an approximately constant carbide volume fraction. The predicted increase therefore supports additional carbide formation, potentially accompanied by coarsening. However, the evolution of the carbide volume fraction alone does not reveal whether this increase results from the nucleation and growth of new carbides, the growth of pre-existing carbides, or a combination of these processes. Comparable overlap between precipitation, growth, and coarsening has been observed experimentally in other carbide-strengthened steels. Thuvander et al. [58] used APT and diffusion modeling to show that V- and Mo-rich MC/$M_2C$ carbides nucleate preferentially at dislocations during ageing of a low-alloyed martensitic steel at 600 °C. With increasing holding time, some particles continued to grow while others had already entered the coarsening stage. Carbide nucleation, growth, and coarsening may therefore overlap rather than occur as three clearly separated stages. Although the material and matrix phase investigated by Thuvander et al. differ from those in the present study, their findings demonstrate that changes in number density, particle size, and morphology do not necessarily represent mutually exclusive stages. The present MatCalc simulations cannot distinguish between homogeneous nucleation in the matrix, heterogeneous nucleation at crystal defects, and direct precipitation at pre-existing carbide–matrix interfaces. The model establishes the available thermodynamic driving force and the expected evolution of the overall carbide population, but it does not independently identify the operative interfacial growth mechanism. This limitation is particularly relevant to the proposed carbide-rim hypothesis. The increase in carbide volume fraction predicted by MatCalc could result from the nucleation and growth of new secondary carbides within the matrix, but it does not exclude additional carbide formation at the interfaces of pre-existing primary carbides. Previous atom-probe studies of high-speed steels have demonstrated that extremely fine secondary carbides and chemically complex primary carbides can coexist within the same microstructure [52–54, 55]. These observations illustrate the presence of multiple carbide populations but do not, by themselves, provide evidence for rim formation. Consequently, the MatCalc simulations support continued carbide precipitation during holding, whereas they cannot determine whether the additional carbide forms as discrete secondary precipitates or contributes to the development of a rim around pre-existing carbides. Confirmation of the carbide-rim mechanism therefore requires spatially resolved experimental evidence of precipitation or compositional evolution at the carbide–matrix interface.

Recent studies also highlight limitations of classical nucleation theory (CNT) at low temperatures, where atomic mobility is restricted. Hutchinson and Brechet [59] proposed a complementary "geometric cluster" model in which nuclei develop from local atomic configurations already present in the material. Ou et al. [60] further demonstrated the role of crystal defects in promoting heterogeneous nucleation in iron. Although these studies do not directly address carbide growth, they emphasize the importance of local atomic structure for phase formation. In the present material, pre-existing carbide–matrix interfaces may provide favorable sites for further growth without requiring the nucleation of separate carbide particles, consistent with the proposed carbide-rim mechanism.

The measured hardness increase is consistent with this combined interpretation. Precipitation during isothermal holding removes carbon and carbide-forming elements from the metastable austenite. The resulting decrease in the carbon concentration of austenite is expected to raise the martensite-start temperature and promote a more complete martensitic transformation during final quenching. The strong dependence of the martensite-start temperature on carbide dissolution and matrix composition has also been demonstrated for PM HS 3-3-4 in a previous work by some of the authors of this article [29]. Even if the retained-austenite fraction of PM HS 3-3-4 is small, a further reduction may contribute to the measured hardness increase. A second contribution can arise from the experimentally observed increase in the volume fraction of hard MC carbides. At the same time, carbon depletion of the austenite reduces the carbon content and thus potentially the intrinsic hardness of the martensite formed during final quenching. The measured net increase in hardness indicates that the effects of the increased MC carbide volume fraction and the potentially larger martensite fraction outweigh this competing reduction in martensite solid-solution strengthening.

While image analysis, thermodynamic calculations, and hardness measurements demonstrate that the carbide population evolves during isothermal holding, they do not reveal whether growth occurs directly at pre-existing carbide interfaces and a chemically distinct rim is formed. Interface-sensitive measurements were therefore required to test the central mechanistic prediction of the hypothesis.

The STEM-EDS investigations provided the first nanoscale indications of gradual compositional transitions across the carbide–matrix interfaces. Such profiles are compatible with the presence of a chemically graded rim. However, the complementary CASINO simulations and MATLAB-based geometric calculations show that comparable broadening may arise from the finite electron-interaction volume and the three-dimensional geometry of the intersected carbides. The apparent STEM-EDS gradients therefore cannot be interpreted as independent proof of chemical grading. A technique with higher spatial resolution is required to distinguish a genuine concentration gradient from geometrical signal broadening.

AES provided complementary surface-sensitive information and revealed gradual compositional transitions at two of the four investigated carbide–matrix interfaces. However, the measurement-point spacing of 19 nm was too large to resolve the expected nanoscale transition reliably. A chemically graded region of comparable or smaller width would be represented by only one or very few measurement points, preventing a reliable determination of its shape and width. Consequently, the AES results indicate possible interfacial modification but cannot establish the presence or dimensions of a chemically graded carbide rim. APT was therefore employed to characterize the carbide–matrix interfaces at substantially higher spatial resolution.

The strongest evidence was obtained by the atom probe tomography results. The directly quenched reference condition exhibits a comparatively sharp carbide–matrix interface with an apparent transition width of approximately 2–3 nm. In contrast, the isothermally held condition shows a gradual transition extending over approximately 10–15 nm. Corresponding gradients are observed for Fe, C, Mo and V, indicating that the difference is not restricted to a single species. The APT results therefore demonstrate that the measured chemical structure of the interface differs between the two heat-treatment conditions.

The quantitative values must nevertheless be interpreted in view of known limitations of APT measurements on carbides. The two heat-treatment conditions were analyzed using different pulsing modes. Voltage pulsing was used for the reference condition, whereas laser pulsing was required for the isothermally held specimens. Liu and Andrén [61] demonstrated that laser pulsing can substantially improve the analysis success rate of fracture-sensitive steels and can provide reliable matrix and precipitate compositions. However, they also identified pulse-mode-dependent effects that must be considered when voltage- and laser-pulsed datasets are compared. Laser-induced heating and surface migration may reduce spatial resolution and broaden an apparent interface.

Part of the difference between the measured transition widths of 2–3 and 10–15 nm could therefore originate from the different APT conditions. Nevertheless, the occurrence of coordinated gradients in several elements, together with the independently measured changes in carbide volume fraction, radius, and circularity, makes a purely instrumental explanation less likely.

The APT-derived transition width of 10–15 nm is smaller than both the 40–80 nm estimate reported for the $M_6C$ carbides investigated by Disch et al. [13] and the approximately 48 nm apparent radial growth increment calculated for the present carbide population. These values are not necessarily contradictory because they quantify different features. The mean-radius difference represents a population-averaged morphological change and may include uniform carbide growth, coarsening, changes in the particle-size distribution, and stereological effects. In contrast, the APT value describes the width over which the local composition changes at one singular carbide–matrix interface. A freshly formed carbide region may therefore be wider than the chemically diffuse part of its outer interface. Conversely, the limited APT sampling volume may have captured an interface that is not representative of the entire carbide population.

The crystallographic nature of the proposed growth requires separate consideration. Pippel et al. [55] observed coherent transitions between nanoscale V-rich carbides and the surrounding matrix in HS 6-5-2-5. Peng et al. [62] likewise identified a preferred crystallographic alignment between MC carbides and the martensitic matrix of the related powder-metallurgical high-speed steel S390, whereas no comparable alignment was detected for $M_6C$ carbides. These observations demonstrate that MC carbides can form crystallographically well-matched interfaces with the matrix in high-speed steel. In such preferred orientations, the atomic arrangements of the carbide and matrix are more compatible at the interface than they would be in an arbitrary orientation. This reduces the lattice mismatch and the associated interfacial strain, thereby promoting the formation of interfaces with comparatively low interfacial energy. Such interfaces can facilitate oriented carbide growth and therefore provide a plausible crystallographic basis for the proposed epitaxial growth mechanism.

In addition to the interface analyses, localized carbon-enriched regions were observed within the martensitic matrix under both heat-treatment conditions. Since these features were detected in both the directly quenched and isothermally held states, they cannot be attributed specifically to the isothermal holding treatment or to the formation of the chemically graded carbide rim. Instead, they most likely reflect the heterogeneous carbon distribution commonly reported in martensitic steels and may result from carbon clustering, segregation to crystal defects, or the earliest stages of carbide precipitation [63, 64].

Taken together, the experimental observations form a consistent sequence of evidence. Image analysis demonstrates that the carbide population evolves during isothermal holding, while the thermodynamic simulations confirm that additional carbide precipitation is energetically favourable. STEM-EDS and AES provide the first indications that chemical modifications occur at the carbide–matrix interface, although neither technique alone allows an unambiguous interpretation. The APT investigations finally resolve the interface with sufficient spatial resolution to reveal a chemically graded transition zone that is absent in the directly quenched reference condition.
The combined evidence therefore provides strong experimental support for the proposed mechanism of epitaxial carbide growth accompanied by the formation of a chemically graded carbide rim. At the same time, the present results do not constitute definitive proof of the mechanism. Rather, they establish a coherent experimental basis showing that the proposed hypothesis is consistent with the observed microstructural evolution and represents the most plausible explanation for the changes occurring during isothermal holding within the transformation gap.

# 5. Conclusion and Outlook

The main findings of the present study can be summarized as follows:

- Isothermal holding within the transformation gap causes measurable microstructural evolution in the powder-metallurgical high-speed steel PM HS 3-3-4. Compared with the directly quenched reference condition, the isothermally held material exhibits a higher carbide volume fraction, increased carbide circularity, an increase in the mean equivalent carbide radius and an increase in hardness.
- The MatCalc simulations predict an increase of the carbide volume fraction during isothermal holding. These results confirm that additional carbide precipitation is thermodynamically and kinetically feasible, although the simulations cannot determine whether precipitation occurs within the matrix, at crystal defects, or directly at pre-existing carbide–matrix interfaces.
- The STEM-EDS and AES investigations provided the first experimental indications of compositional changes at the carbide–matrix interfaces, thereby offering important initial evidence for the proposed formation of a chemically graded carbide rim.
- APT reveals a comparatively sharp carbide–matrix interface with an apparent transition width of approximately 2–3 nm in the directly quenched reference condition. In contrast, the isothermally held condition exhibits coordinated concentration gradients of Fe, C, Mo, and V extending over approximately 10–15 nm. This difference provides strong evidence that the chemical structure of the carbide–matrix interface changes during isothermal holding.

Taken together, the changes in carbide volume fraction, size, circularity, hardness, and interfacial composition are consistent with interface-assisted growth of pre-existing MC carbides and the formation of a chemically graded carbide rim. The results therefore provide strong support for the proposed mechanism but do not constitute definitive proof of epitaxial carbide growth, as the crystallographic relationship across the investigated interfaces was not determined.

Future work should focus on extending the experimental evidence by investigating additional carbide–matrix interfaces and different holding conditions to further assess the generality of the proposed mechanism. Direct confirmation of epitaxial growth requires crystallographic characterization of the carbide core, the proposed rim, and the surrounding matrix. Site-specific TEM specimens containing complete carbide–rim–matrix interfaces could be investigated using high-resolution TEM, nanobeam or precession electron diffraction, and orientation mapping by 4D-STEM. These techniques could determine whether the proposed rim possesses the same crystal structure and orientation as the pre-existing carbide and whether a defined crystallographic relationship exists with the matrix. Moreover, correlating the evolution of the chemically graded carbide rim with the resulting mechanical properties may enable the targeted design of heat-treatment strategies for optimizing the performance of powder metallurgical high-speed steels.

# Acknowledgments

The authors gratefully acknowledge financial support from the Deutsche Forschungsgemeinschaft (DFG, German Research Foundation) under Project-ID 532214230.